\documentclass[a4paper]{llncs}
\usepackage{url, color, graphicx, amsmath, colordvi, subfigure, epsfig}
\usepackage{capt-of}
\usepackage{verbatim}
\usepackage{algorithm}
\usepackage[noend]{algpseudocode}

\newcommand\T{\rule{0pt}{2.6ex}}
\newcommand\B{\rule[-1.2ex]{0pt}{1pt}}

\begin{document}
\title{Phoenix: An Epidemic Approach to Time Reconstruction}
\author{Jayant Gupchup$^\ast$ Douglas Carlson$^\ast$  R\u{a}zvan Mus\u aloiu-E.$^\ast$ Alex Szalay$^\ddagger$ Andreas Terzis$^\ast$}
\institute{
Computer Science Department$^\ast$ \quad Physics and Astronomy Department$^\ddagger$ \\
Johns Hopkins University \\ 
{\texttt{\{gupchup,carlsondc,razvanm,terzis\}@jhu.edu}$^\ast$ \quad
  \texttt{szalay@jhu.edu}$^\ddagger$ \\
}
}

\maketitle

\begin{abstract}

Harsh deployment environments and uncertain run-time conditions create
numerous challenges for postmortem time reconstruction methods. For
example, motes often reboot and thus lose their clock state,
considering that the majority of mote platforms lack a real-time
clock. While existing time reconstruction methods for long-term data
gathering networks rely on a persistent basestation for assigning
global timestamps to measurements, the basestation may be unavailable
due to hardware and software faults. We present {\em Phoenix}, a novel
offline algorithm for reconstructing global timestamps that is robust
to frequent mote reboots and does not require a persistent global time
source. This independence sets Phoenix apart from the majority of time
reconstruction algorithms which assume that such a source is always
available. Motes in Phoenix exchange their time-related state with
their neighbors, establishing a chain of transitive temporal
relationships to one or more motes with references to the global
time. These relationships allow Phoenix to reconstruct the measurement
timeline for each mote. Results from simulations and a deployment
indicate that Phoenix can achieve timing accuracy up to 6 ppm for
99\% of the collected measurements. Phoenix is able to maintain this
performance for periods that last for months without a persistent
global time source. To achieve this level of performance for the
targeted environmental monitoring application, Phoenix requires an
additional space overhead of 4\% and an additional duty cycle of
0.2\%.

\end{abstract}

\section{Introduction}
\label{sec:intro}

Wireless sensor networks have been used recently to understand spatiotemporal
phenomena in environmental studies \cite{ucb-hab,TPS+05}. The data these
networks collect are scientifically useful only if the collected measurements
have corresponding, accurate global timestamps. The desired level of accuracy
in this context is in the order of milliseconds to seconds. In order to reduce
complexity of the code running on the mote, it is more efficient to record
sensor measurements using the mote's local time frame and perform a postmortem
reconstruction to translate them to global time.

Each mote's clock (referred to as local clock henceforth)
monotonically increases and resets to zero upon reboot. A naive
postmortem time reconstruction scheme collects $\langle local,
global\rangle$ pairs during a mote's lifetime, using a global clock
source (typically, an NTP-synchronized PC).  These pairs (also
referred to as ``anchor points'') are then used to translate the
collected measurements to the global time frame by estimating the
motes' clock skew and offset.  We note that this methodology is
unnecessary for architectures such as Fleck, which host a
battery-backed on-board real-time clock (RTC) \cite{FLECK}. However,
many commonly-used platforms such as Telos, Mica2, MicaZ, and IRIS
(among others) lack an on-board RTC.

In the absence of reboots, naive time reconstruction strategies
perform well.  However, in practice, motes reboot due to low battery
power, high moisture, and software defects. Even worse, when motes
experience these problems, they may remain completely inactive for
non-deterministic periods of time. Measurements collected during
periods which lack $\langle local, global\rangle$ anchors (due to
rapid reboots and/or basestation absence) are difficult or impossible
to accurately reconstruct. Such situations are not uncommon based on
our deployment experiences and those reported by others \cite{ALJL+06}.

In this work, we devise a novel time reconstruction strategy, {\em Phoenix},
that is robust to random mote reboots and intermittent connection to the global
clock source. Each mote periodically listens for its neighbors to broadcast
their local clock values.  These $\langle local, neighbor\rangle$ anchors are
stored on the mote's flash.  The system assumes that one or more motes can
periodically obtain global time references, and they store these $\langle
local, global\rangle$ anchors in their flash. When the basestation collects the
data from these motes, an offline procedure converts the measurements
timestamped using the motes' local clocks to the global time  by using the
transitive relationships between the local clocks and  global time.

The offline nature of Phoenix has two advantages: {\bf (a)} it reduces the
complexity of the software running on the mote, and {\bf (b)} it avoids the
overhead associated with executing a continuous synchronization protocol. We
demonstrate that Phoenix can reconstruct global timestamps accurately (within
seconds) and achieve low $(< 1 \%)$ data losses in the presence of random mote
reboots even when months pass without access to a global clock source.

\section{Motivation}
\label{sec:mot}

We claim that the problem of rebooting motes is a practical aspect of
real deployments that has a high impact on environmental monitoring
applications. We also quantify the frequency and impact of reboots in
a long-term deployment. We begin by understanding why mote reboots
complicate postmortem time reconstruction.

\subsection{Postmortem Timestamp Reconstruction}
\label{subsec:ptr}

The relationship between a mote's local clock, $LTS$, and the global
clock, $GTS$, can be modeled with a simple linear relation: $GTS =
\alpha \times LTS + \beta$, where $\alpha$ represents the mote's skew
and $\beta$ represents the intercept (global time when the mote reset
its clock) \cite{Sallai+EWSN06}.  This conversion from the local clock
to global clock holds as long as the mote's local clock monotonically
increases at a constant rate. We refer to this monotonically
increasing period as a {\em segment}. When a mote reboots and starts a
new segment, one needs to re-estimate the fit parameters.  If a mote
reboots multiple times while it is out of contact with the global
clock source, estimating $\beta$ for these segments is difficult.
While data-driven treatments have proven useful for recovering
temporal integrity, they cannot replace accurate timestamping
solutions~\cite{SUN+EWSN+09,TIME+IPSN+09}. Instead, time
reconstruction techniques need to be robust to mote reboots and not
require a persistent global time source.

\begin{figure}[t]
\begin{center}
  \includegraphics[scale=0.5]{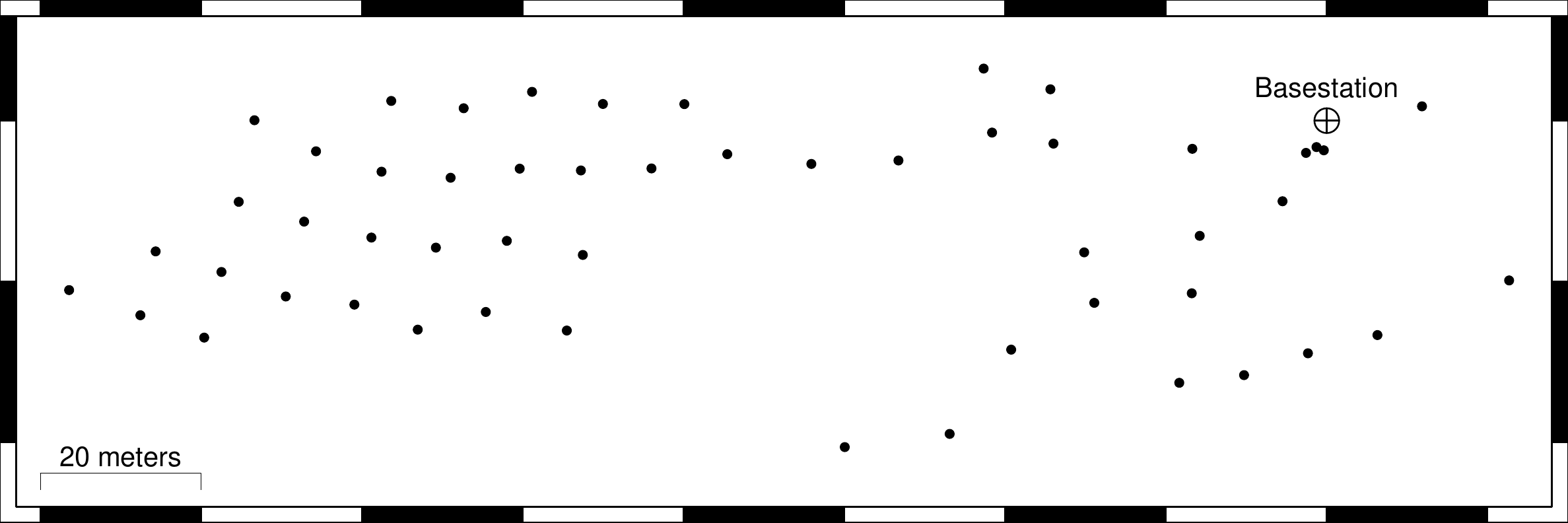}
  \caption {The 53-mote ``Cub Hill'' topology, located in an urban
    forest northeast of Baltimore, Maryland.}
  \label{fig:topo}
\end{center}
\end{figure}

\subsection{Case Studies}

We present two cases which illustrate the deployment problems that
Phoenix intends to address.  The first is an account of lessons
learned from a year-long deployment of 53 motes.  The second is a
result of recent advances in solar-powered sensor networks.

\paragraph{Software Reboots.}

We present ``Cub Hill'', an urban forest deployment of $53$
motes that has been active since July 2008 (Figure
\ref{fig:topo}). Sensing motes collect measurements every
10 minutes to study the impact of land use on soil
conditions. The basestation uses the Koala protocol to collect
data from these motes every six hours~\cite{ipsn:koala:l}. We use TelosB
motes driven by  $19$ Ah, $3.6$ V batteries.

\begin{figure}[t]
\begin{center}
  \includegraphics[scale=0.5]{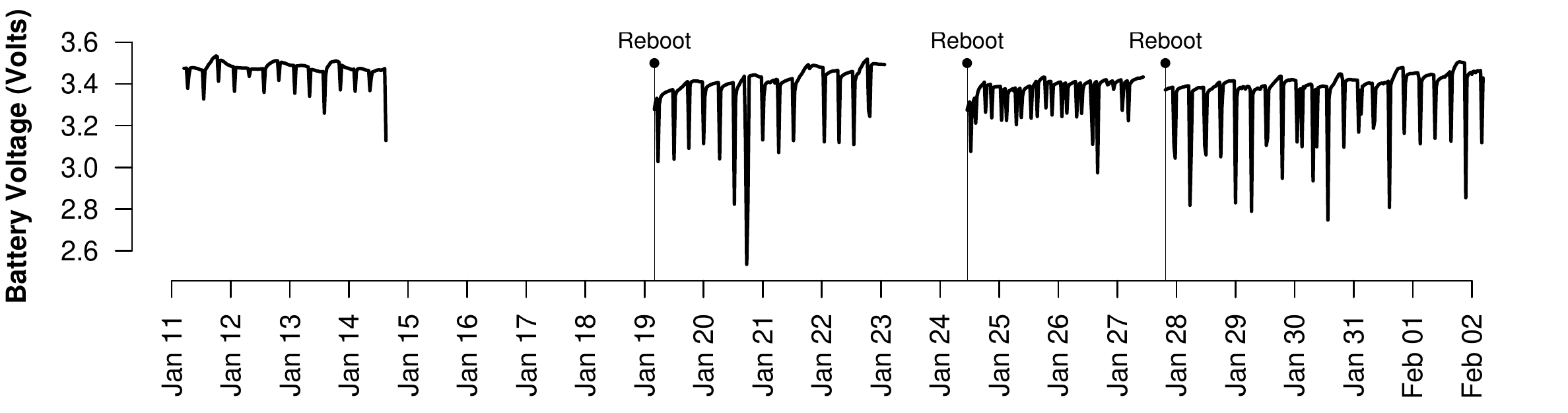}
  \caption {An example of a mote rebooting due to low battery voltage
(no watchdog timer in use). The sharp downward spikes correspond to
gateway downloads (every six hours). Gaps in the series
are periods where the mote was completely inoperative. }
  \label{fig:reboot_bv}
\end{center}
\end{figure}

We noticed that motes with low battery levels and/or
high internal moisture levels suffered from periodic reboots. As an
example, Figure \ref{fig:reboot_bv} shows the battery voltage of a
mote that rebooted thrice in one month. Despite their instability,
many of these motes were able to continue collecting measurements for
extended periods of time. 

\begin{figure}[t]
\begin{center}
  \includegraphics[scale=0.5]{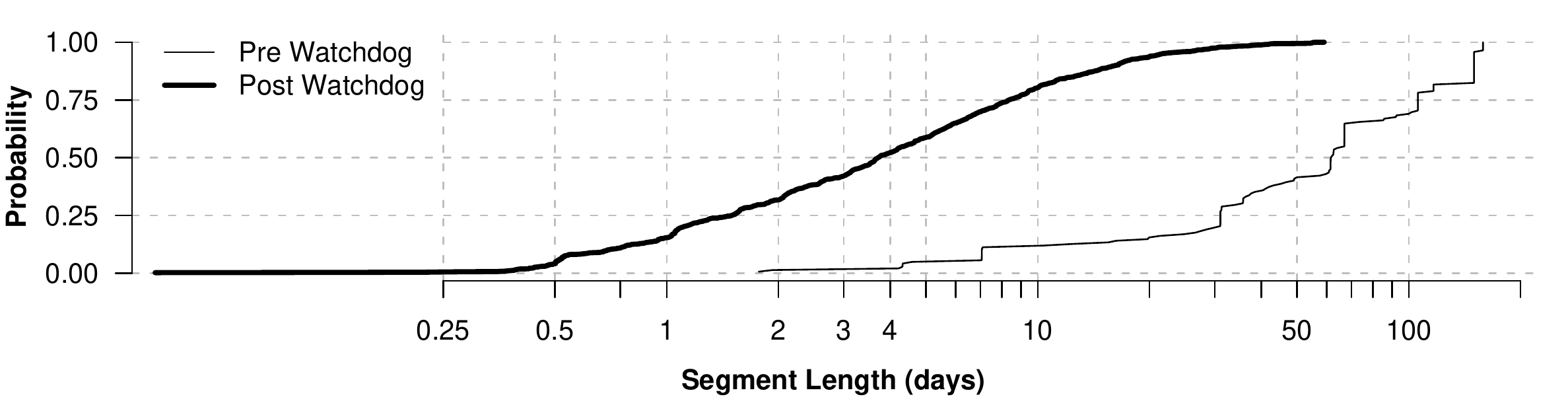}
  \caption {The distribution of the segment lengths before
   and after adding the watchdog timer to the mote software.}
  \label{fig:seglen}
\end{center}
\end{figure}

Following a major network expansion, a software fault appeared which
caused nodes to ``freeze''. Unable to reproduce this
behavior in a controlled environment, we employed the MSP430's
Watchdog Timer to reboot motes that enter this state~\cite{MSP430}.
While this prevented motes from completely failing, it also shortened
the median length of the period between reboots from more than 50 days
to only four days, as Figure \ref{fig:seglen} shows.

\paragraph{Solar Powered Sensor Networks.} A number of research groups
have demonstrated the use of solar energy as a means of powering
environmental monitoring sensor networks
\cite{LHAS+INFOCOM2007,TANEJA+IPSN+08}.  In such architectures, a mote
can run out of power during cloudy days or at night.  Motes naturally
reboot in such architectures, and data losses are unavoidable due to
the lack of energy. It is unclear how one can achieve temporal
reliability without a persistent basestation or an on-board RTC. To
the best of our knowledge, no one has addressed the issue of temporal
integrity in solar-powered sensor networks.  Yang et al. employ a
model in which data collection happens without a persistent
basestation~\cite{SOLARSTORE+MOBISYS+09}.  The data upload takes place
infrequently and opportunistically. Hard-to-predict reboot behavior is
common to these systems.  Furthermore, we note that even though there
is very little information about the rate of reboots in such
architectures, it is clear that such systems are susceptible to
inaccurate timestamp assignments.

\subsection{Impact}
\label{sec:impact}

We evaluate the impact of mote reboots on the Cub Hill deployment using
our existing time reconstruction methodology. 

The basestation records an anchor point each time it downloads data
from a mote. Motes that are poorly connected to the basestation may
remain out of contact for several download rounds before connectivity
improves and they can transfer their data. When motes reboot at a rate
faster than the frequency with which the basestation contacts them,
there exist periods which lack enough information to accurately
reconstruct their measurement timestamps.

Upon acquiring the anchor points, the measurements are converted
from their local clock to the global clock at the basestation. We
employ our previously proposed algorithm, Robust Global Timestamp
Reconstruction algorithm (referred to as RGTR), for this
purpose~\cite{SUN+EWSN+09}. We note that in order to estimate the fit
parameters ($\alpha$, $\beta$) for the segments, RGTR requires at
least two anchor points. Depending on the accuracy requirements, one can
assume that the skew ($\alpha$) is stable per mote for small
segments. Using this assumption, at least one anchor point is needed
to estimate the $\beta$ for any given segment, provided that $\alpha$
has been estimated accurately for the mote.

\begin{figure}[t] 
 \centering 
 \subfigure[The fraction of measurements that were assigned timestamps.]
{
\label{sfig:recon} 
\includegraphics[scale=0.5]{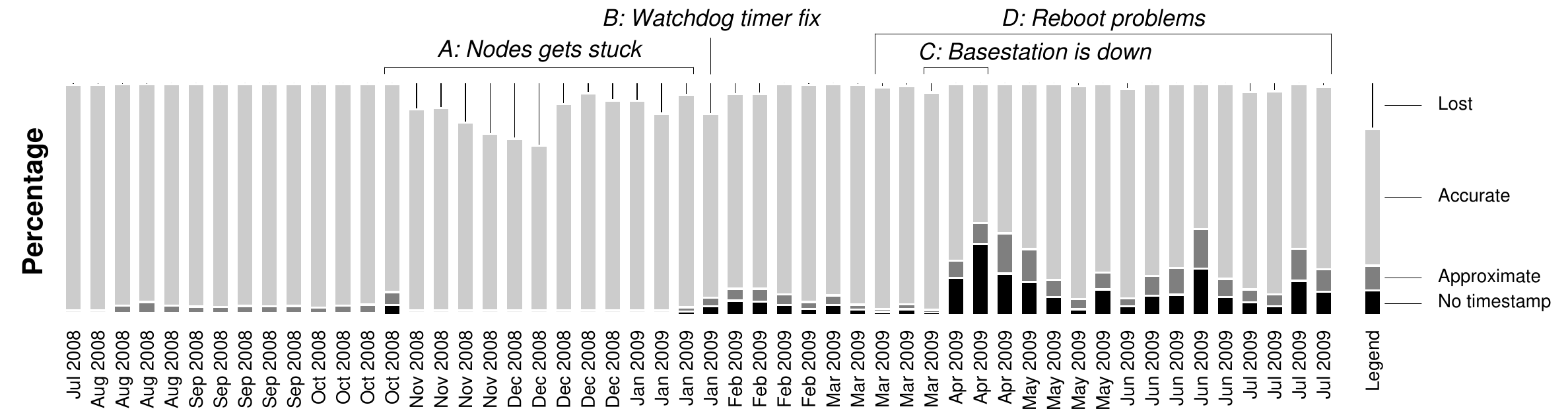}
}
\subfigure[An example of the impact of estimating $\beta$ incorrectly
  when using approximate methods. Data from one of the motes
    (represented with the dark line) that rebooted multiple times
    between Jun. 22 and Jun. 25. During this period, the mote was
    out of sync with the rest (shown in gray) due to inaccurate
    $\beta$ estimates] 
{
  \label{sfig:appx} \includegraphics[scale=0.5]{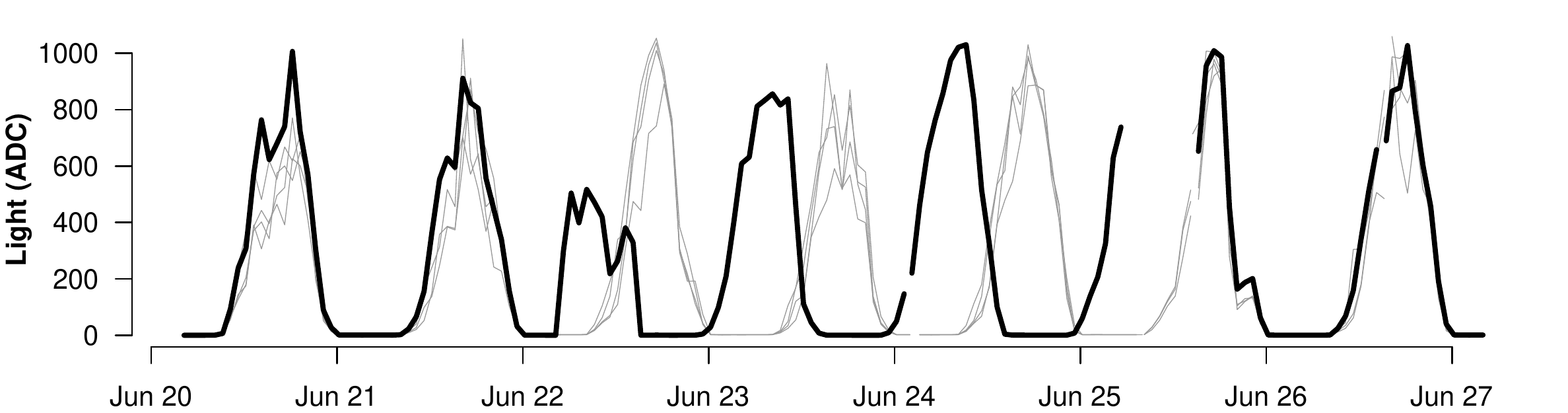}
}
\caption{Impact of time reconstruction methodology using the RGTR
  algorithm.}
\label{fig:impact}
\end{figure}

Figure \ref{sfig:recon} demonstrates the impact of mote reboots on
time reconstruction for the Cub Hill deployment. During
  period A, motes were prone to freezing (and thus stopped sampling),
leading to a decrease in the total data collected. At point B, the
addition of the watchdog timer caused the total data collected to
return to its previous level.  However, due to the increased frequency
of reboots, a larger portion of the samples could not be assigned a
global timestamp (exacerbated by the absence of the base station
during period C).

For segments where no anchor points were collected, we assumed
that node reboots are instantaneous. However, this assumption does not
always hold (see Figure~\ref{fig:reboot_bv}) and leads to a small
fraction of misaligned measurements. Figure~\ref{sfig:appx} presents
an example of this misalignment.  One node (shown in bold) rebooted
multiple times and could not reach the basestation during its active
periods. The assumption of instantaneous reboots led to inaccurate
$\beta$ estimates.

\section{Solution}
\label{sec:soln}

Phoenix is a postmortem time reconstruction algorithm for
motes operating without in-network time synchronization.
It consists of two stages.

\subsection{In-Network Anchor Collection}
\label{subsec:anchors}

Each mote operates solely with respect to its own local clock.   
A new segment (uniquely identified by $\langle moteid, reboot~count\rangle$)
begins whenever a mote reboots: each segment starts at a different time
and may run at a different rate. Our architecture
assumes that there is at least one mote in the network that can
periodically obtain references from an accurate global time
source. This is done to establish the global reference points needed
by Phoenix. This source may be absent for long periods of time (see
Section \ref{sec:results}).  The global time source can be any
reliable source (a mote equipped with a GPS receiver, NTP-synced
basestation, etc). Without loss of generality, we assume that the
network contains a mote connected to GPS device and a basestation that
collects data infrequently\footnote{Note that the basestation collects
data  {\em only} and it does not provide a time source, unless specified
otherwise.}.

All motes (including the GPS-connected mote) broadcast their local
clock and reboot-count values every $T_{beacon}$ seconds. Each receiving mote
stores this information (along with its own local clock and reboot
counter) in flash to form anchor records. The format of these records
is $\langle moteid_r, rc_r,lc_r,moteid_s,rc_s,lc_s\rangle$; where $rc,
lc, r, $ and $s$ refer to the reboot counter, local clock, receiver and
sender respectively. Periodically, motes turn on their radios and
listen for broadcasts in order to anchor their time frame to those of
their neighbors. Each mote tries to collect this information from its
neighbors after every reboot and after every $T_{wakeup}$ seconds ($
\gg T_{beacon}$).  The intuition behind selecting this strategy is as
follows. The reboot time determines the $\beta$ parameter. The
earliest opportunity to extract this information is immediately after
a reboot.  To get a good estimate of the skew, one would like to
collect multiple anchors that are well distributed in time. Thus,
$T_{wakeup}$ is a parameter that governs how far to spread out
anchor collections. In the case of a GPS mote, the $moteid_r, rc_r$
and $moteid_s,rc_s$ are identical, and $lc_r, lc_s$ represent the
local and global time respectively.

The basestation periodically downloads these anchors along with the
measurements.  This information is then used to assign global
timestamps to the collected measurements using Algorithm
\ref{algo:phoenix}. If the rate of reboots is known, the anchor
collection frequency can be fixed conservatively to collect enough
anchors between reboots. One could also employ an
adaptive strategy by collecting more anchors when the segment is small
and reverting to a larger $T_{wakeup}$ when an adequate number of
anchors have been collected. It is advantageous for a mote to attempt
to collect anchors from a small set of neighbors (to minimize
storage), but this requires a mote to have some way of identifying the
most useful segments for anchoring (see Section \ref{sec:results}).

\begin{algorithm}[t]
\caption{Phoenix}
\label{clockrecon-algo}
\begin{algorithmic}\scriptsize
\Ensure  
\State $a,b$ : alpha and beta for local-local fits; 
\State $P$ : parent segment; $\Pi$ : Ancestor segments
\Statex
\Procedure{Phoenix}{$AP$} 
\For {each $(i,j)$ in \Call{Keys}{$AP$}}
\Comment{All unique segment pairs in $AP$}
\State $LF_{a,b,\chi,df}(i,j) \gets$ \Call{Llse}{$AP(i,j)$}   
\Comment{Compute the local-local fits}
\EndFor

\For {each $s$ $\in$ $S$}
\Comment{Set of all unique segments}
\State $GF_{\alpha,\beta,P,\Pi,\chi,df}(s) \gets
(\emptyset,\emptyset,\emptyset,s,\chi_{MAX},\emptyset)$  
\Comment{Initialize global fits}
\EndFor

\For {each $g$ $\in$ $G$}
\Comment{All segments anchored to GTS}
\State \Call{InitGTSNodes}{$g,LF,GF$}
\State \Call{Enqueue}{$Q,g$}
\Comment{Add all the GTS nodes to the queue}
\EndFor

\While {\Call{NotEmpty}{$Q$}}
\State $q \gets $ \Call{Dequeue}{$Q$}
\State $C \gets $ \Call{NeighborAnchors}{$q$}
\For {each $c \in C$}
\State $T_{\alpha,\beta,P,\Pi,\chi,df}(c) \gets $\Call{GlobalFit}{$c,q,GF,LF$}
\If {(\Call{UpdateFit}{$c,T,GF$})}
\Comment {Check for a better fit}
\State \Call{Enqueue}{$C$}
\EndIf 
\EndFor
\EndWhile
\State return $GF$
\EndProcedure

\Statex


\Procedure{InitGTSNodes}{$g,LF,GF$}

\State $GF(g) \gets 
(LF_a(g,g'), LF_b(g,g'), \emptyset, g, LF_\chi(g,g'), LF_{df}(g,g'))$
\Comment{$g'$ is GTS, $g$ is LTS}
\EndProcedure

\Statex


\Procedure{GlobalFit}{$c,q,GF,LF$}
\If {$q > c$}
\Comment  Smaller segment is the independent variable
\State $\alpha_{new} \gets GF_\alpha(q) * LF_{a}(q,c)$
\State $\beta_{new} \gets GF_\alpha(q)*LF_{b}(q,c) + GF_\beta(q)$
\Else
\State $\alpha_{new} \gets GF_\alpha(q) / LF_a(q,c)$
\State $\beta_{new} \gets GF_\alpha(q) - \alpha_{new} * LF_b(q,c)$
\EndIf

\State $\chi \gets \frac{GF_{df}(q)*GF_\chi(q) +
  LF_{df}(q,c)*LF_\chi(q,c)}{GF_{df}(q) + LF_{df}(q,c)}$
\Comment Compute the weighted $GOF$ metric.
\State $df \gets GF_{df}(q) + LF_{df}(q,c)$
\State return $(\alpha_{new}, \beta_{new}, q, \{c \cup GF_\Pi(q)\},\chi, df)$
\Comment Update parent/ancestors
\EndProcedure

\Statex

\Procedure{UpdateFit}{$c,T,GF$}
\If {$c \in T_\Pi(c)$}
\Comment {Check for cycles}
\State return false
\EndIf

\If {$T_{\chi}(c) < GF_{\chi}(c)$}
\State $GF_{\alpha,\beta,P,\Pi,\chi,df}(c) \gets T_{\alpha,\beta,P,\Pi,\chi,df}(c)$
\State return true
\Else
\State return false
\EndIf
\EndProcedure 


\end{algorithmic}
\label{algo:phoenix}
\end{algorithm}


\subsection {Offline Timestamp Reconstruction}
\label{subsec:phoenix}

The Phoenix algorithm is intuitively simple. We will outline it in text and
draw attention to a few important details. For a more complete treatment,
please refer to the pseudocode in Algorithm~\ref{clockrecon-algo}. Phoenix
accepts as input the collection of all anchor points $AP$ (both $\langle local,
neighbor\rangle$ and $\langle local, global\rangle$). It then employs a
least-square linear regression to extract the relationships between the local
clocks of the segments that have anchored to each other ($LF$, for Local Fit).
In addition to $LF_a(i,j)$ (slope), $LF_b(i,j)$ (intercept), Phoenix also
obtains a goodness-of-fit ($GOF$) metric, $LF_\chi(i,j)$ (unbiased estimate of
the variance of the residuals) and $LF_{df}$ (degrees of freedom).  For
segments which have global references, Phoenix stores this as $GF$ (for Global
Fit).

The algorithm then initializes a queue with all of the segments which have
direct anchors to the global clock. It dequeues the first element $q$ and
examines each segment $c$ that has anchored to it. Phoenix uses the transitive
relationship between $GF(q)$ and $LF(q,c)$ to produce a global fit $T(c)$ which
associates segment $c$ to the global clock through segment $q$. If $T_\chi(c)$
is lower than the previous value for $GF_\chi(c)$ (and using $q$ would not
create a cycle in the path used to reach the global clock), the algorithm
replaces $GF(c)$ with $T(c)$, and places $c$ in the queue. When the queue is
empty, no segments have ``routes'' to the global clock which have a better
goodness-of-fit than the ones which have been previously established. At this
point, the algorithm terminates.

The selection of paths from an arbitrary  segment to a segment with global time
references can be thought of as a shortest-path problem (each segment
represents a vertex and the fit between the two segments is an edge). The $GOF$
metric represents the edge weight.   The running time complexity of the
implementation of Phoenix was validated experimentally by varying the
deployment lifetime (thereby varying number of segments). The runtime was found
to increase slower than the square of the number of segments.

\section{Evaluation}
\label{sec:results}

We evaluate the effect of varying several key parameters in Phoenix using 
both simulated and real datasets. We begin by describing our simulator.

\subsection{Simulator}
\label{subsec:sim}

Our goal is to minimize the data loss in long-term deployments. Hence,
we fix the simulation period to be one year. We also assume that the
basestation is not persistently present and does not provide a time
source to the network. The network contains one global clock source (a
GPS mote) that is susceptible to failures. The main components of the
simulator are described below. The default values for the simulator
are based on empirical data obtained from the one year long Cub Hill
deployment. \\

\noindent \textbf{Clock Skew: }The clock skew for each segment is drawn from a
uniformly distributed random variable between 40 ppm and 70 ppm. Burri
et al. report this value to be between 30 and 50 ppm at room
temperature\footnote{We ignore the well-studied
temperature effects on the quartz crystal. For a more complete
treatment on the temperature dependence, refer
to \cite{QUARTZ,CRYSTAL+TEMP}.} \cite{Dozer07}.

\noindent \textbf{Segment Model: }We use the non-parametric
segment-length model based on the Cub Hill deployment after the watchdog
timer fix (Figure \ref{fig:seglen}). Additionally, after a reboot, we
allowed the mote to stay inactive for a period that is randomly drawn
between zero and four hours with a probability given by $p_{down}=0.2$
. The GPS mote's behavior follows the same model.

\noindent \textbf{Communication Model:} The total end-to-end
communication delay for receiving anchor packets is drawn uniformly
between 5 and 15 milliseconds. This time includes the interrupt
handling, transmission, reception and propagation delays.  To model
the packet reception rate (PRR), we use the log-distance path loss
model as described in \cite{RAPPAPORT,LOG-DISTANCE} with parameters:
$(P_{r}(d_0), \eta, \sigma, d_0) = (-59.28, 2.04, 6.28, 2.0 m)$.

\noindent \textbf{Topology: }The Cub Hill topology was used as the
basis for all simulations.

\noindent \textbf{Event Frequencies: }Motes recorded a 26-byte sample
every 10 minutes. They beacon their local clock values with an
interval of $T_{beacon}$. They stay up after every reboot and
periodically after an interval of $T_{wakeup}$ to collect these
broadcasts. While up, they keep their radios on for a maximum of
$T_{listen}$. The GPS mote collects $\langle local,global\rangle$
anchors with a rate of $T_{sync}$. By default,
$T_{beacon}$, $T_{wakeup}$, $T_{listen}$ and $T_{sync}$ were set to
30~s, 6~h, 30~s and 6~h respectively.



\noindent \textbf{Maximum Anchorable Segments: }To minimize the space
overhead in storing anchors, we limit the number of segments that can
be used for anchoring purposes. At any given time, a mote can only 
store anchors for up to $NUMSEG$ segments. The default $NUMSEG$ value is
set to four. Motes stop listening early once they collect $NUMSEG$
anchors in a single interval.

\noindent \textbf{Eviction Policy: }Since segments end and links between motes
change over time, obsolete or rarely-heard segments need to be evicted from the
set of $NUMSEG$ segments for which a mote listens.  The timeout for evicting
stale entries is set to $3 \times T_{wakeup}$. We evaluated three different
strategies for selecting replacements for evicted segments.  First-come,
first-served (FCFS) accepts the first segment that is heard when a vacancy
exists. RAND keeps track of the previous segments that were heard and selects a
new segment to anchor with at random. Longest local clock (LLC) keeps track of
the local clock values of the segments that are heard and selects the segment
that has the highest local clock. FCFS was chosen as the default.


\begin{figure}[t] \centering 
\subfigure[The effect of a missing global clock source on accuracy.]
{\label{sfig:nogps_ppm} \includegraphics[width=0.95\textwidth]{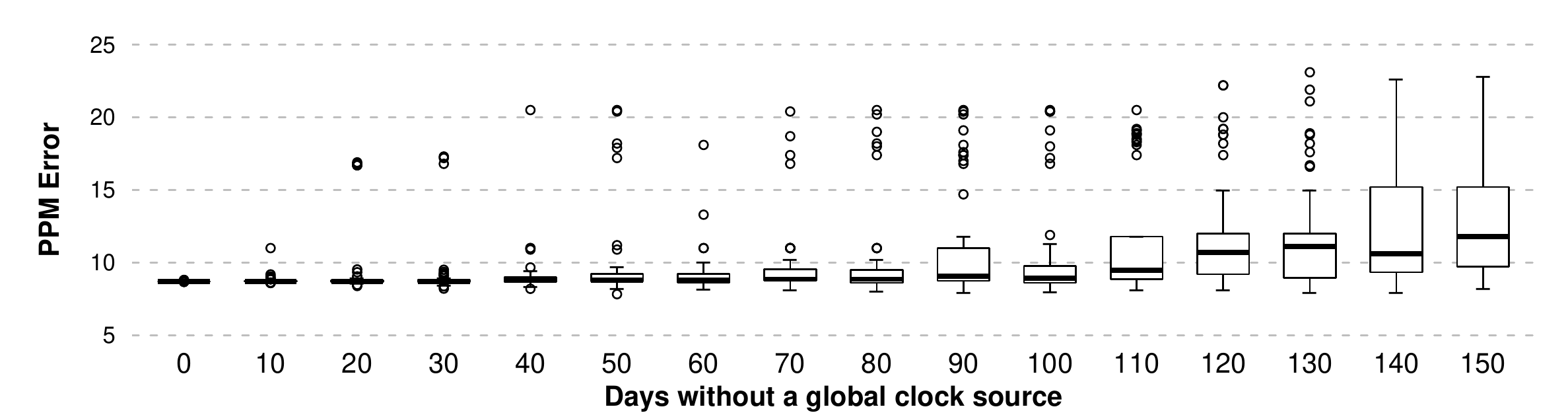}}
\subfigure[The impact of $T_{wakeup}$ on data loss.]
{\label{sfig:twake_dloss} \includegraphics[width=0.485\textwidth]{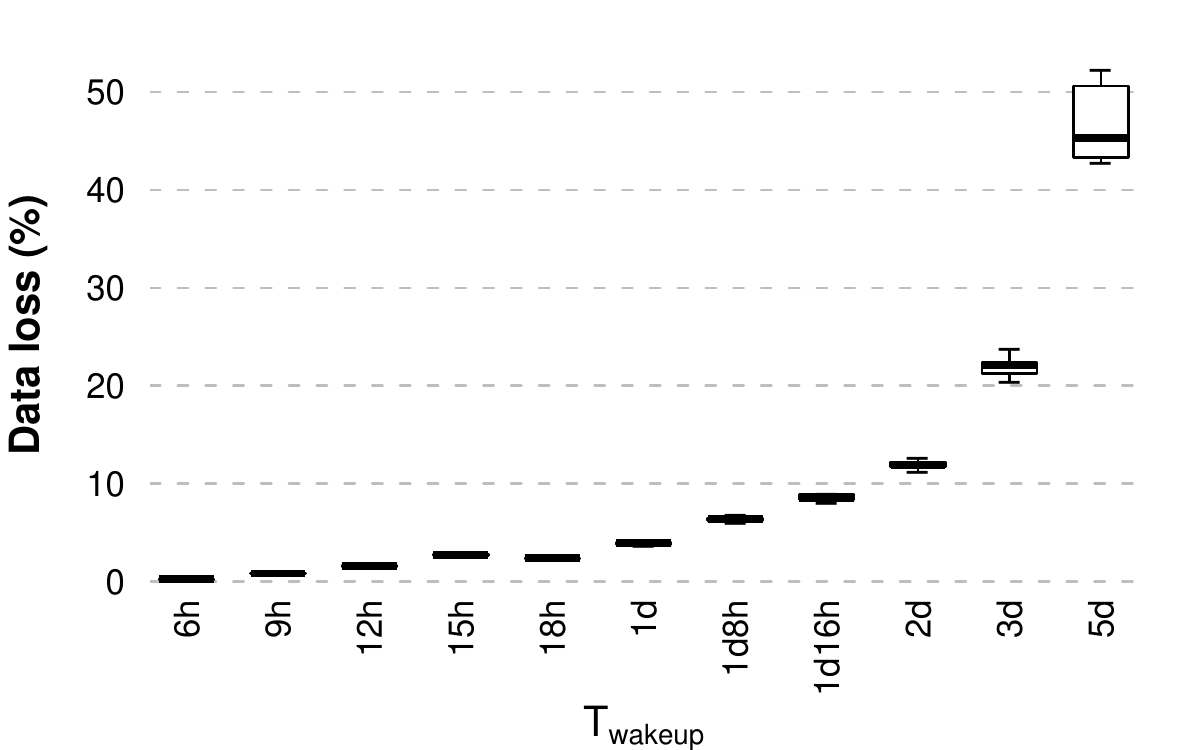}}
\subfigure[Robustness to bad anchors.]
{\label{sfig:corr_apt} \includegraphics[width=0.485\textwidth]{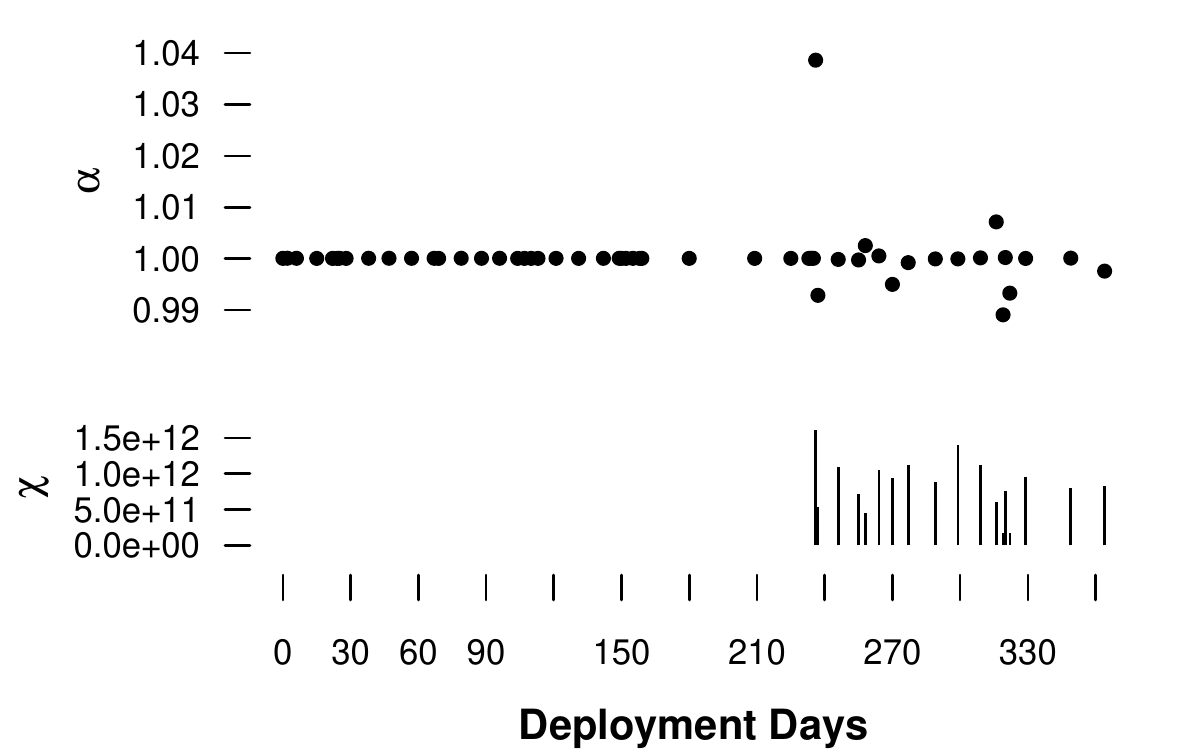}}
  \caption{Evaluation of Phoenix in simulation. In (c), faults were
injected to GPS anchors after day 237.  Figure shows the $\alpha$ and
$\chi$ values for the GPS mote for the entire period.}
  \label{fig:phoenix_eval}
\end{figure}

\subsection{Evaluation metrics}
\label{subsec:met}


\begin{description}
\item[ Data loss (DL):] The fraction of data that cannot be assigned
  any timestamps, expressed as a
  percentage. 

\item[ PPM Error:] The average error (in parts per million) for the
  assigned timestamps. PPM error is $\frac{|t^\prime-t|}{t_\delta} \times 10^6$,
  where $t$ is the true timestamp of the
  measurement, $t^\prime$ is the assigned timestamp, and $t_\delta$
  denotes the elapsed time since the start of the segment in terms of
  the real clock.

\item[ Space overhead:] The fraction of space that is used for storing
  anchors relative to the total space
  used, expressed as a percentage.

\item[ Duty cycle:] The fraction of time the radio was kept on for
anchor collection and beaconing, expressed as a percentage.

\end{description}

\subsection{Simulation Experiments}

\paragraph{Dependence on Global Clock Source: } We studied the effect of the
global clock's absence on data loss.  We assume that the network contains one
GPS mote that serves as the global clock source and it is inoperative for a
specified amount of time. In order to avoid bias, we randomly selected the
starting point of this period and varied the GPS down time from 0 to 150 days
in steps of 10.  Figure \ref{sfig:nogps_ppm} shows the effect on the
reconstruction using 60 independent runs. The accuracy decreases as the number
of days without GPS increases, but we note that this decrease is tolerable for
our target applications. The data loss stayed relatively stable at $0.21\%$,
even when the global clock source is absent for as long as 5 months.  We note
that in a densely connected network, the number of paths between any two
segments is combinatorial, and hence, the probability of finding a usable path
is very high\footnote{One can estimate the probability for finding a usable
path using Warshall's algorithm \cite{Cormen}. The input to this algorithm
would be a connectivity matrix where the entries represent the anchoring
probabilities of the neighbor segments.}. The variance of the error increased
with the length of the gateway's absence.

\paragraph{Dependence on Wake-up Interval: } Figures \ref{sfig:twake_dloss}
show the effect of varying wake-up rate on data loss.  As expected, data loss
increases as the rate of anchor collection decreases. This curve is strongly
related to the segment model: if collections are less frequent than reboots,
many segments will fail to collect enough anchors to be reconstructed. 

\paragraph{Robustness: } We studied the effect of faulty global clock
references on time reconstruction. Noise from a normal distribution
($\mu=$ 60 min., $\sigma=$10 min.) was added to the global references
for a period of 128 days. Figure \ref{sfig:corr_apt} shows the alpha
and $\chi$ values for the GPS mote during the entire simulation
period. One can also notice the correlation between high $\chi$
values and $\alpha$ values that deviate from 1.0 in Figure
\ref{sfig:corr_apt}. These faults did not change the data loss rate.
The faults increased the PPM error from $4.03$ to $16.5$. Although these
faults decreased accuracy, this decrease is
extremely small in comparison to the magnitude of the injected errors
and within the targeted accuracy requirements. Phoenix extracted paths
which were least affected by these faults by using the $\chi$ metric. 

\begin{figure}[t]
\centering 
\subfigure[Space overhead in storing anchors as a function of $NUMSEG$.]
{\label{sfig:e_space} \includegraphics[scale=0.49]{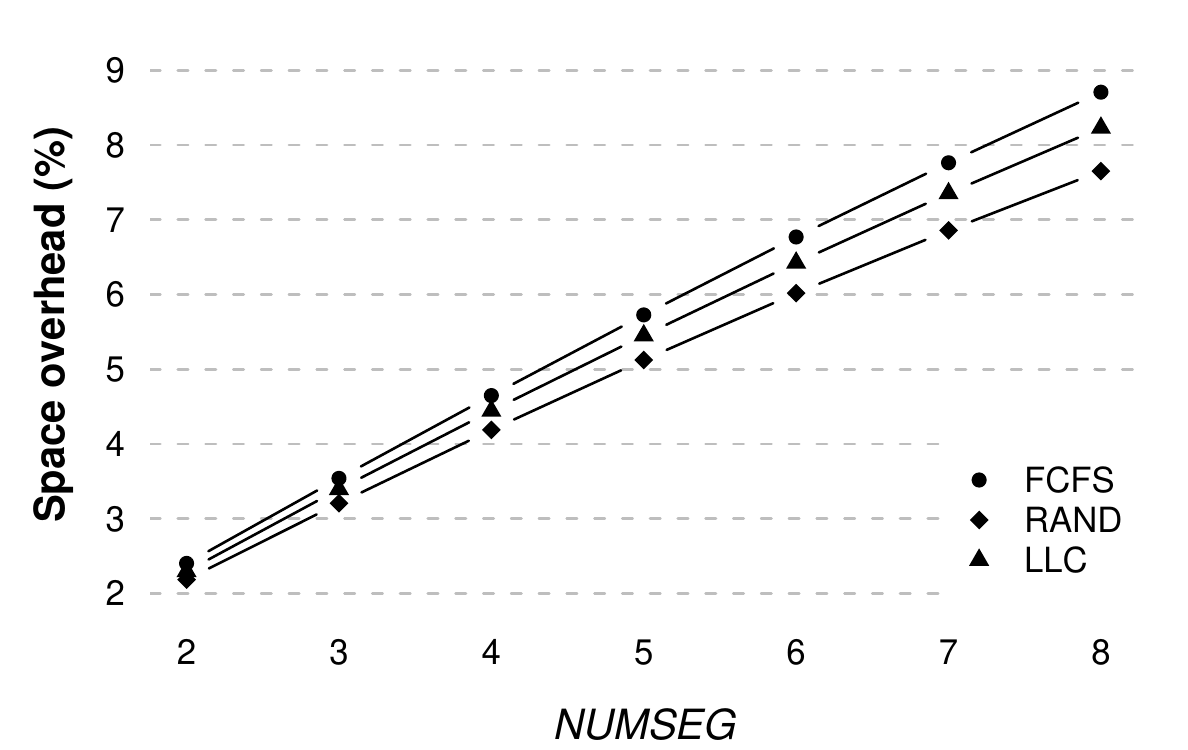}}
\subfigure[Duty cycle as a function of $NUMSEG$.]
{\label{sfig:e_dc} \includegraphics[scale=0.49]{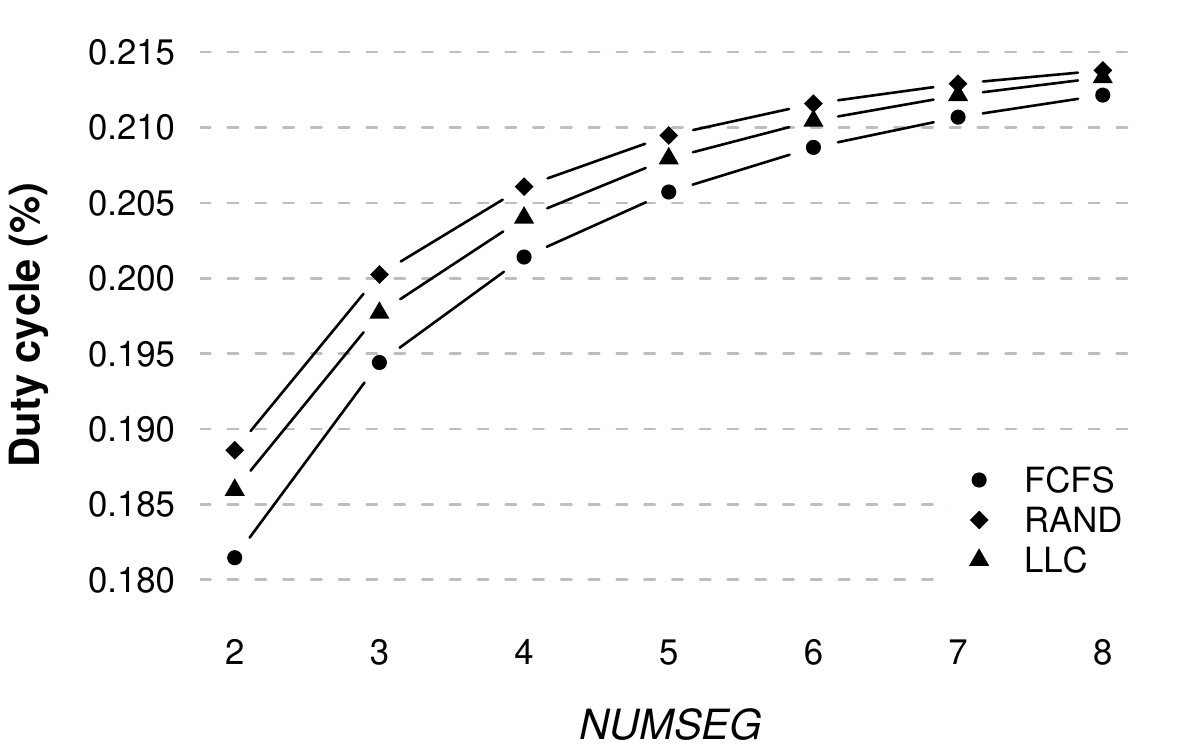}}
\subfigure[Data loss as a function of $NUMSEG$.]
{\label{sfig:e_dloss} \includegraphics[width=0.485\textwidth]{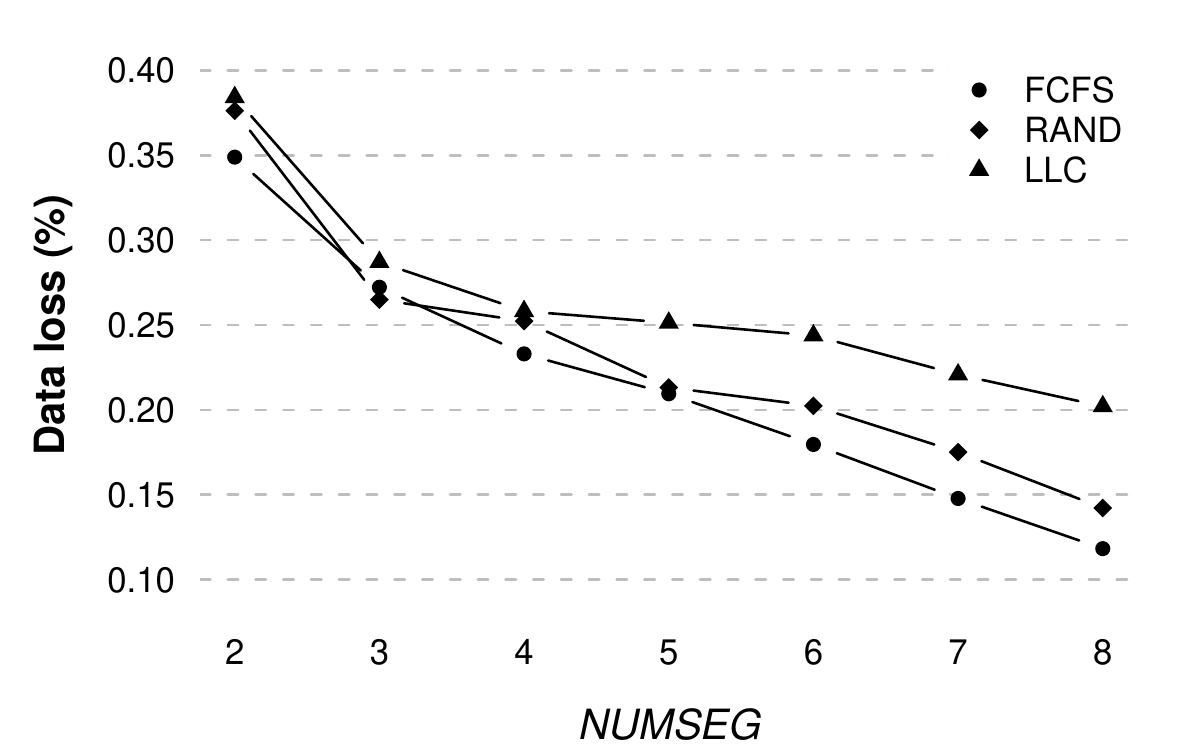}}
\subfigure[Effect of varying node density on data loss with no eviction policy.]
{\label{sfig:d_dloss} \includegraphics[width=0.485\textwidth]{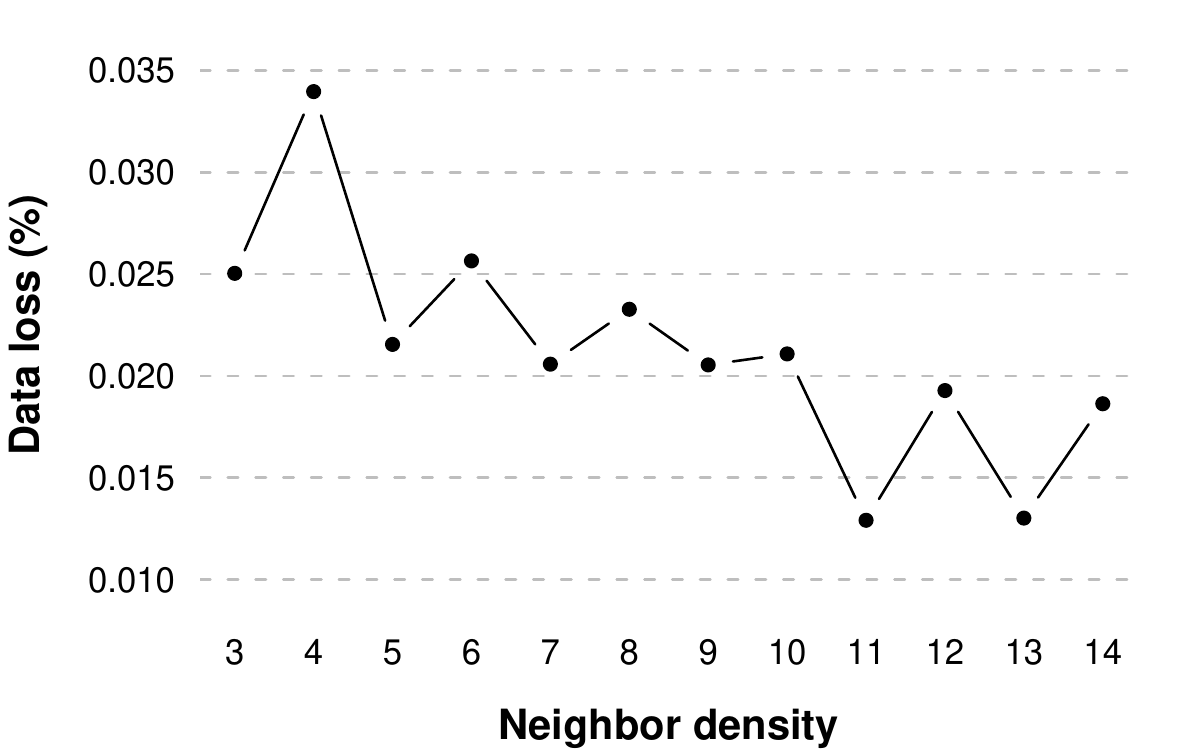}}
 \caption{Effect of $NUMSEG$ on different eviction policies.}
  \label{fig:evict_policy}
\end{figure}

\paragraph{Effect of eviction and $NUMSEG$:} We studied the effect of $NUMSEG$
on space, duty cycle, and data loss. The space overhead increases linearly with
$NUMSEG$ (Figure \ref{sfig:e_space}).  The impact on duty cycle\footnote{Note
that the duty cycle that we are referring to does not consider the
communication costs during data downloads. Reducing the storage requirements
would reduce the communication costs when the basestation collects data.} was
quite low (Figure \ref{sfig:e_dc}). A constant duty cycle penalty of 0.075\% is
incurred due to the beaconing messages sent every 30~s~\cite{ipsn:koala:l}. At
low values of $NUMSEG$, motes are able to switch off their radios early (once
they have heard announcements from segments they have anchored with), while at
higher values, they need to stay on for the entire $T_{listen}$ period.
Increasing $NUMSEG$ decreases data loss, because motes have a better chance of
collecting good segments to anchor with. We found that the FCFS eviction policy
outperforms LLC and RAND. We found no significant differences in the PPM error
results as we vary $NUMSEG$, and hence, we do not report those results here.

\paragraph{Neighbor Density: } In this experiment, we removed
links from the Cub Hill topology until we obtained the desired neighbor
density.  At every step, we ensured that the network was fully
connected. We did not find any significant impact on performance as
the average number of neighbors was decreased.  In this experiment,
the radios were kept on for the entire $T_{listen}$ period, and no
eviction policy was employed. This was done to compare the performance
at each density level at the same duty cycle. Figure
\ref{sfig:d_dloss} presents our findings.

\subsection{Deployment - I}

We deployed a network (referred to as the ``Olin'' network) of 19 motes
arranged in a grid topology in an urban
forest near the Johns Hopkins University campus in Baltimore, MD.
Anchors were collected for the entire period of 21 days using
the methodology described in Section \ref{subsec:anchors}. The basestation
collected data from these motes once every four hours and the
NTP-corrected clock of the basestation was used as a reliable global
clock source.  The motes rebooted every 5.7 days
on average, resulting in a total of 62 segments. The maximum segment
length was 19 days and the minimum was two hours.  


\paragraph{Perceived Ground Truth: } It is very difficult to establish
absolute ground truth in field experiments. Instead, we 
establish a synthetic ground truth by reconstructing timestamps using
all the global anchors obtained from the basestation\footnote{Note
that every time a mote contacts the basestation, we obtain a global
anchor for that mote.}. We record the $\alpha$ and $\beta$ values
for each segment and use these values as ground truth. Because we
downloaded data every four hours we obtained enough global anchors from
the motes to be confident with the derived ground truth estimates.

\paragraph{Emulating GPS node and Basestation Failure: } In order to
emulate a GPS mote, we selected a single mote (referred to as G-mote)
that was one hop away from the basestation. We used the G-mote's global
anchors obtained from the basestation as though they were taken using
a GPS device. We ignored all other global anchors obtained from other
motes. Furthermore, to emulate the absence of the basestation for $N$
days, we discarded all the anchors taken by the G-mote during that
$N$-day long period. We tested for values of $N$ from one to eighteen.

\begin{table} [t]
\caption{Phoenix accuracy using the Olin dataset as a function of
  the number of days that the basestation was unavailable.} 
\begin{center}
\begin{tabular}{l|*{9}{@{\extracolsep{1em}}r@{.\extracolsep{0pt}}l}}
Error\textbackslash Days & 2 & 4 & 6 & 8 & 10 & 12 & 14 & 16 & 18\B\\
\hline 
$\alpha_{med}$ (ppm) &  1&73 &  1&73 &  1&85 &  1&70 &  1&96 &
2&20 & 4&36 & 5&47 & 5&93\T\\
$\alpha_{std}$ (ppm) & 3&41 & 3&40 & 3&40 & 3&39  & 3&30 &
3&26 & 3&17 & 3&00 & 3&00 \B\\
\hline
$\beta_{med}$ (s)    &  0&88 & 0&88 & 0&91 & 0&94 & 1&16 &  
1&55 & 4&52 & 6&02 & 6&44 \T\\
$\beta_{std}$ (s)    &  0&58 &  0&57 & 0&58 & 0&57  & 0&65 &
0&91 & 2&43 & 3&11 & 3&45\B\\
\hline

\end{tabular}
\label{tab:error}
\end{center}
\end{table}

\begin{figure}[t]
\centering 
\subfigure[The CDF of $\alpha$ estimates on the Olin deployment]
{\label{sfig:o_alpha} \includegraphics[scale=0.485]{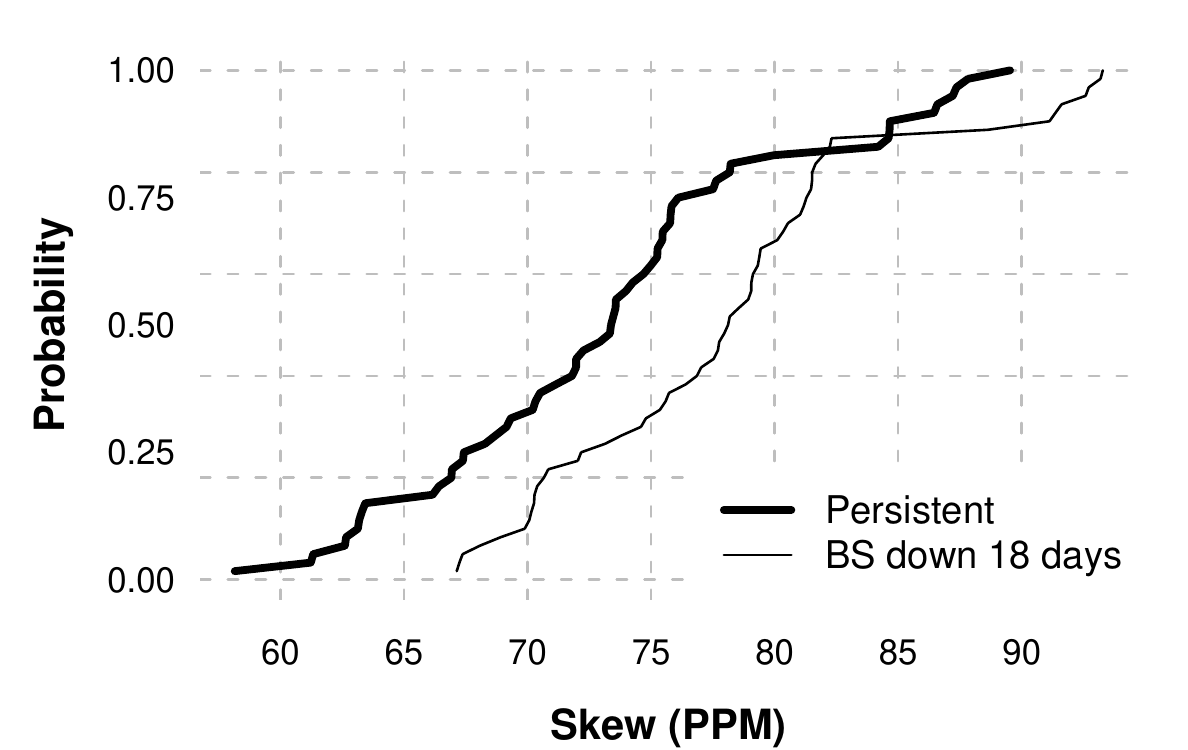}}
\subfigure[Data loss using RGTR. Data loss from Phoenix was $< 0.06\% $.]
{\label{sfig:o_dl} \includegraphics[scale=0.485]{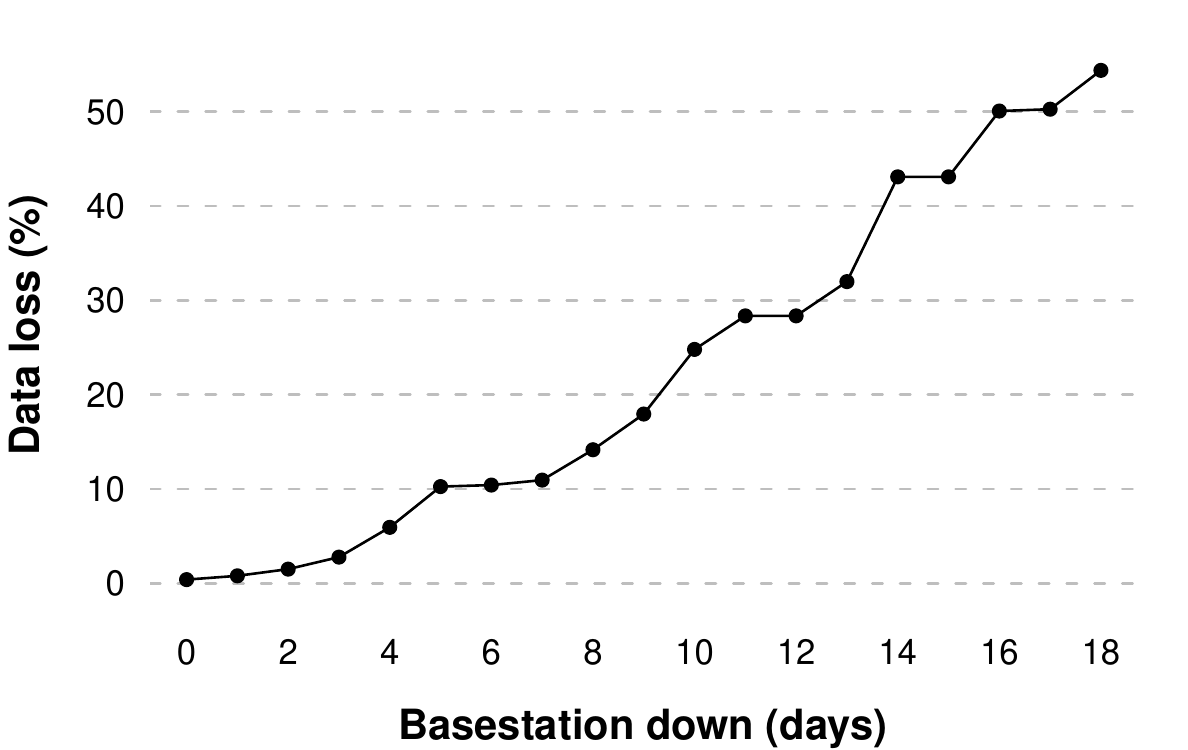}}
\caption{The stability of the $\alpha$ estimates using Phoenix and
  the data loss using RGTR in comparison to Phoenix.}
  \label{fig:olin}
\end{figure}

\paragraph{Phoenix Accuracy: } After simulating the basestation
failure, we reconstruct the timestamps by applying Phoenix using only
the $\langle local, neighbor\rangle$ anchors, and global anchors
available from the G-mote. This provides us with another set of
$\alpha$ and $\beta$ estimates for each of the segments. We compare
these estimates with the ground truth estimates (pair-wise
comparison). In order to provide a deeper insight, we decompose the
average PPM error metric into its constituent components - $\alpha$
and $\beta$ errors. Furthermore, we report the median and standard
deviation of these $\alpha$ and $\beta$ errors. Table \ref{tab:error}
reports the results of these experiments. We found that the
median $\alpha$ error stayed as low as 5.9 ppm, while the median
$\beta$ error stayed as low as 6.4 s for $N=$18.  In general,
$\alpha_{med}$, $\beta_{med}$ and $\beta_{std}$ increased as $N$
increased and $\alpha_{std}$ stayed relatively consistent for
different values of $N$. The stability of the $\alpha$ estimates using
Phoenix with $N=0$ and $N=18$ is shown in Figure
\ref{sfig:o_alpha}. The CDF shows that median skew was found to be
around 75 ppm and the two curves track each other closely.

\paragraph{Data Loss: } The data loss using Phoenix was found to be as
low as $0.055\%$ when $N$ was 18 days. In comparison, we found that
there was significant data loss when the timestamps were reconstructed
using RGTR. Figure \ref{sfig:o_dl} shows the data
losses for different values of $N$. The figure does not report the
Phoenix data loss as we found it to be $0.055\%$ irrespective of
$N$. This demonstrates that Phoenix is able to reconstruct more than
$99\%$ of the data even when motes reboot frequently and the
basestation is unavailable for days. We note that in comparison to
Phoenix, RGTR does not incur any additional storage and duty
cycle overheads as anchors are recorded at the basestation
directly as part of the data downloads.
 
\subsection{Deployment - II}
The second deployment (termed Brazil) was at the Nucleo Santa Virginia research
station in the Atlantic coastal rain forest near Sao Paolo, Brazil
\cite{k2+carlson}. The goal of this deployment was to collect data to improve
atmospheric  micro-front models. $52$ nodes were deployed for a total of $35$
days and $5,418,074$ data points were produced during this campaign. The site
could not host a persistent basestation. Instead, researchers would download
data every alternate data using a laptop that served as a temporary mobile
basestation. The basestation was running a linux VM over windows 7 - our
download protocol required a linux installation.

\paragraph{Deployment Setup: } Two GPS receivers were built on two motes and
these were to serve as the global clock source. These motes would advertise
their local clock values for others to anchor with and would also periodically
store the (local,GPS) timestamps on their flash - this is in addition to storing
time state announcements from other motes. However, due to the lithium 
battery shipping problems, these GPS motes were unavailable until 22 days
into the deployment. Due to these problems, we had to use the 
laptop's VM clock as the global clock source for the first 22 days. After
the batteries arrived, we found out that one of the GPS receivers did not
work.

\begin{figure}[t]
\begin{center}
  \includegraphics[scale=0.95]{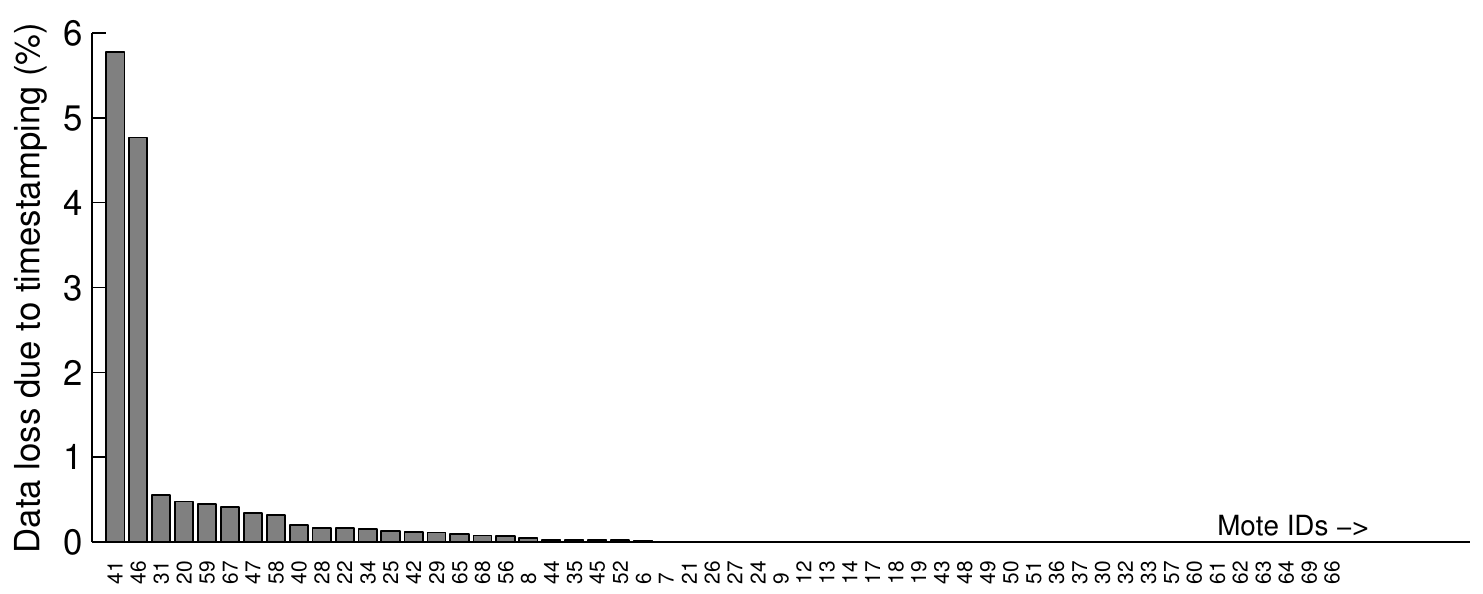}
  \caption {Data loss due to timestamping for the motes in the Brazil deployment}
  \label{fig:brazil_yield}
\end{center}
\end{figure}

\begin{figure}[t]
\begin{center}
  \includegraphics[scale=0.95]{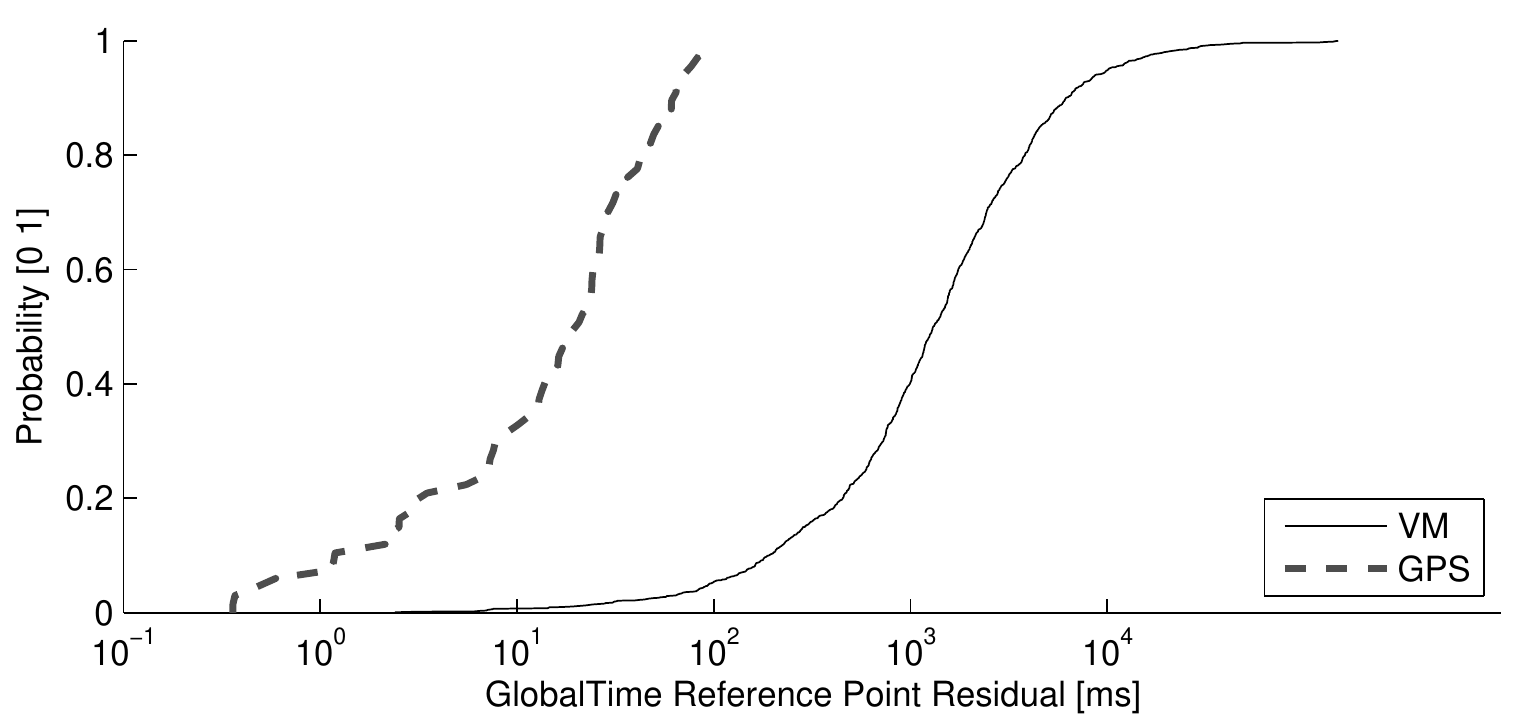}
  \caption {Residuals of fits with global time references for the VM clock
  and the GPS.}
  \label{fig:brazil_residuals}
\end{center}
\end{figure}

\paragraph{Experiences: }When we looked at the temperature time
series plots of the reconstructed data for the first few days, we found a few
motes ``shifted'' and ``out-of-sync'' from one another. The peaks and troughs
in the temperature seemed lagged at a few sensors. The motes initially started
in-sync and then gradually went out of sync. This indicated to us that some of
the motes had poor estimates of $\alpha$. On further investigation, we realized
that the VM clock was highly unstable and this lead to poor reconstruction. 

Our only hope was to then rely on using the GPS anchors, available from
day $22$ to day $35$ collected by the one working GPS mote. Even though we did not
intend things to go this way, this situation was exactly what phoenix was designed for
- tolerance to a missing global clock source for extended periods of time.

\paragraph{Results: }Using the GPS anchors, Phoenix was able to timestamp
$99.7\%$ of all the data that was collected. The data loss due to timestamping
for all the motes in the Brazil deployment is shown in Figure \ref{fig:brazil_yield}.
Other than mote $41$ and $46$, more than $70\%$ of the motes have  
less than $0.1\%$ of timestamping data loss.

The accuracy of these timestamps is difficult to report since ground truth was
not available to us. Nonetheless, we can compare the relative quality of the
two global clock sources. The CDF of the residuals for the fits obtained using
the VM clock and the GPS clock is shown in Figure \ref{fig:brazil_residuals}.
Note that low residuals indicate a good linear fit between mote clocks and the
reference clock. By looking at the distribution in Figure
\ref{fig:brazil_residuals}, the median residual for GPS is almost two orders of
magnitude lower than the VM clock. One can also notice the long tail (high
errors) in the distribution of the VM residuals. The effect of temperature on
the mote clock and non-deterministic delays in the GPS interrupt handling
account for variation in the GPS residuals. An obvious, but often overlooked,
take away from this experience is to ensure that the global clock source is
trustworthy and accurate - just having one is not good enough.

\section{Related Work}
\label{sec:relwork}

Assignment of timestamps in sensor networks falls under two broad
categories. Strict clock synchronization aims at ensuring that all the
mote clocks are synchronized to the same clock source. Flooding Time
Synchronization Protocol (FTSP, \cite{FTSP}), Reference Broadcast
Synchronization (RBS, \cite{EGE02}), and the Timing-sync Protocol for
Sensor Networks \cite{GKS03} are examples of this approach. These
systems are typically used in applications such as target tracking and
alarm detection which require strong real-time guarantees of
reporting events. 
The second
category is known as postmortem time reconstruction and it is mostly
used due to its simplicity. While strict synchronization is
appropriate for applications where there are specific events of
interest that need to be reported, postmortem reconstruction is
well-suited for applications where there is a continuous data stream
and every measurement requires an accurate timestamp.

Phoenix falls under the second class of methods. The idea of using
linear regression to translate local timestamps to global timestamps
was first introduced by Werner-Allen et al. in a deployment that was
aimed at studying active volcanoes \cite{ALJL+06}. This work, however,
does not consider the impact caused by rebooting motes and basestation
failures from a time reconstruction perspective. More recently,
researchers have proposed data-driven methods for recovering temporal
integrity \cite{SUN+EWSN+09,TIME+IPSN+09}. Lukac et al. use a
model for microseism propagation to time-correct the data collected by
their seismic sensors. Although data-driven methods have proved useful
for recovering temporal integrity, they are not a solution for
accurate timestamping.

Routing integrated time synchronization protocol (RITS,
\cite{Sallai+EWSN06}) spans these categories. Each mote along the path
(to the basestation) transforms the time of the reported event from
the preceding mote's time frame, ending with an accurate global
timestamp at the basestation. RITS does not consider the problem of
mote reboots, and is designed for target tracking applications. The
problem of mote reboots have been reported by a number of research
groups.  Chang et al. report that nodes rebooted every other day due
to an unstable power source \cite{CHANG08}, whereas Dutta et
al. employed the watchdog timer to reboot nodes due to software faults
\cite{Dutta06trio}. Allen et al. report an average node uptime of
$69\%$ \cite{ALJL+06}.  More recently, Chen et al. advocate {\em Neutron},
a solution that detects system violations and recovers from them
without having to reboot the mote \cite{Chen+SIGOPS09}. They advocate
the notion of preserving ``precious'' states such as the time
synchronization state. Nevertheless, Neutron cannot prevent all mote
reboots and therefore Phoenix is still necessary.

\section{Conclusions}
\label{sec:concl}

In this paper we investigate the challenges facing existing
postmortem time reconstruction methodologies due to basestation
failures, frequent random mote reboots, and the absence of on-board
RTC sources. We present our time reconstruction experiences based on a
year-long deployment and motivate the need for robust time
reconstruction architectures that minimize data losses due to the
challenges we experienced.

Phoenix is an offline time reconstruction algorithm that assigns
timestamps to measurements collected using each mote's local
clock. One or more motes have references to a global time source. All
motes broadcast their time-related state and periodically record the
broadcasts of their neighbors. If a few mote segments are able to map
their local measurements to the global time frame, this information
can then be 
used to assign global timestamps to the measurements collected by
their neighbors and so on. This epidemic-like spread of global
information makes Phoenix robust to random mote reboots and
basestation failures. We found that in practice there are more than
enough possible ways to obtain good fits for
the vast majority of data segments.

Results obtained from simulated datasets showed that Phoenix is able
to timestamp more than 99\% of measurements with an accuracy up to 6
ppm in the presence of frequent random mote reboots. It is able to
maintain this performance even when there is no global clock
information available for months.  The duty-cycle and space overheads
were found to be as low as 0.2\% and 4\% respectively. We validated
these results using a 21 day-long real deployment and were able to
reconstruct timestamps in the order of seconds.

In the future, we will investigate using other metrics for determining
edge weights and their impact on the quality of the time
reconstruction. Moreover, we will explore adaptive techniques for
determining the anchor collection frequency. Finally, we will derive
theoretical guarantees on the accuracy of Phoenix, which can be used
to allow for fine-grained tradeoffs between reconstruction quality and
overhead.

\section*{Acknowledgments}
\label{sec:ack}

We thank Prabal Dutta, Jay Taneja and the anonymous reviewers for
their comments that helped us to improve the paper's
presentation. This research was supported in part by NSF grants
DBI-0754782 and CNS-0720730. Any opinions, finding, conclusions or
recommendations expressed in this publication are those of the authors
and do not represent the policy or position of the NSF.

\begin{small}
\bibliographystyle{abbrv} 
\bibliography{phoenix}

\end{small}

\end{document}